\documentclass[twocolumn,prb,showpacs,showkeys,preprintnumbers,amsmath,amssymb]{revtex4}
\usepackage{graphicx}
\usepackage{dcolumn}
\usepackage{bm}

\begin{document}
\preprint{Ren et. al., submitted to PRB}

\title{Upper critical field of aligned Na$_{x}$(H$_{3}$O)$_{z}$CoO$_{2}\cdot y$H$_{2}$O superconductor by magnetization measurement}

\author {Zhi Ren,$^{1}$ Chunmu Feng,$^{2}$ Zhu'an
Xu,$^{1}$ Minghu Fang,$^{1}$ Guanghan Cao$^{1}$\footnote[1]{To whom correspondence should be addressed
(ghcao@zju.edu.cn)}}

\affiliation{$^{1}$Department of Physics, Zhejiang University, Hangzhou 310027, People's Republic of China}

\affiliation{$^{2}$Test and Analysis Center, Zhejiang University, Hangzhou 310027, People's Republic of China}

\date{\today}

\begin{abstract}
We present magnetization measurements for the upper critical field $H_{c2}$ on aligned
Na$_{x}$(H$_{3}$O)$_{z}$CoO$_{2}\cdot y$H$_{2}$O superconductors with magnetic field parallel to CoO$_2$ layers.
The temperature-dependent part of the normal-state magnetization $M(H,T)$ can be well scaled with Brillouin
function, suggesting the existence of impurity moments. By subtracting the contribution of the magnetic impurity
moments, a well defined onset of superconducting diamagnetism allows the in-field superconducting transition
temperature $T_{c}$($H$) accurately determined. No superconducting transition was observed down to 1.9 K under
an applied field of 8 Tesla. The result suggests that the $H^{\|ab}_{c2}(0)$ value is just within the Pauli
paramagnetic limit, supporting spin-singlet superconductivity in the cobalt oxyhydrate superconductor.
Additionally, the upward curvature near $T_{c}$ in the $H^{\|ab}_{c2}-T$ phase diagram was confirmed to be
intrinsic. Possible origin of the anomaly was discussed.
\end{abstract}


\maketitle
\section{\label{sec:level1}INTRODUCTION}

Superconductivity in Na$_{x}$CoO$_{2}\cdot y$H$_{2}$O \cite{1} has attracted much attention in recent years. The
new layered superconductor exhibits both similarities and differences with the well-known cuprate
superconductors, shedding light on the challenging issue of understanding high-$T_{c}$ superconductivity.
\cite{2} At present, evidences of unconventional superconductivity in this material have been indicated or
implied by experimental researches \cite{3,4,5,6,7,8,9} as well as by theoretical studies. \cite{10,11,12}
However, some of the basic questions on the superconductivity remain open. For example, an unprecedented number
of proposals have been suggested for the superconducting pairing symmetry. \cite{13} Obviously, clarification of
the spin state of Cooper pairs is crucial to this problem.

Upper critical field $H_{c2}$ is among the most important superconducting parameters since it not only directly
correlates with the superconducting coherence length $\xi$, the size of Cooper pairs, but gives a clue to the
pairing symmetry. However, the determination of $H_{c2}(0)$ in Na$_{x}$CoO$_{2}\cdot y$H$_{2}$O superconductor
remains controversial. Resistivity,\cite{14} heat capacity \cite{9,15} and $^{59}$Co-NMR \cite{16} measurements
suggested that the $H_{c2}(0)$ value was within the Pauli paramagnetic limit\cite{17,18} $H_{P}$ $\approx$ 8.3
T, implying spin-singlet state for the Cooper pairs. In contrast, a much larger $H_{c2}(0)$ value, far exceeding
the Pauli paramagnetic limit, was inferred from magnetization measurements,\cite{4,8} which favors spin-triplet
superconductivity. Another issue is the observation of an abrupt slope change of $H_{c2}$ (or upward curvature)
on the $H_{c2}-T$ curve near $T_{c}$,\cite{8,9} which was explained in terms of field-induced pairing symmetry
transition.\cite{8,19} Since the anomalous curvature was observed on randomly oriented polycrystalline sample,
the anomaly could be an extrinsic phenomenon due to the anisotropy of $H_{c2}$.\cite{9}

In order to clarify these issues we carried out magnetization measurements for $H_{c2}$ on \emph{aligned} cobalt
oxyhydrate superconductors. In principle, $H_{c2}-T$ phase diagram can be obtained by determining the
superconducting transition temperature under different magnetic field, $i.e.$, $T_{c}$($H$). In the case of the
cobalt oxyhydrate superconductor, however, determination of $T_{c}$($H$) becomes difficult especially at high
field due to the susceptibility upturn in the normal state.\cite{4} We found that the temperature-dependent part
of normal-state magnetization can be well scaled in terms of Brillouin function, suggesting that the
susceptibility upturn above $T_{c}$ arises from impurity moments rather than spin fluctuation. The in-field
superconducting transition temperature $T_{c}$($H$) was then accurately determined by subtracting the
contribution of magnetic impurity. Consequently the estimated $H_{c2}(0)$ was found to be just within the Pauli
paramagnetic limit. Besides, the upward curvature near $T_{c}$ in the $H_{c2}-T$ phase diagram was confirmed to
be intrinsic.

\section{\label{sec:level1}EXPERIMENT}

Polycrystalline samples of the cobalt oxyhydrate superconductor were synthesized through three steps as
described previously.\cite{5,20} First, parent compound Na$_{0.7}$CoO$_{2}$ was prepared by a solid-state
reaction from high purity Na$_{2}$CO$_{3}$ and Co$_{3}$O$_{4}$ powders. Second, the sodium was adequately
deintercalated by employing Br$_{2}$/CH$_{3}$CN as an oxidizing agent. Third, the resulted Na$_{x}$CoO$_{2}$
samples with $x\sim$ 0.33 were then hydrated at room temperature in saturated NaCl solution for 10 days. X-ray
diffraction confirmed that the obtained samples were single phase. The chemical composition of the sample was
determined by the techniques described previously.\cite{20} The sample used in present study has the chemical
composition of Na$_{0.33}$(H$_{3}$O)$_{0.02}$CoO$_{2}\cdot 1.4$H$_{2}$O.

We made use of the large anisotropy in magnetization \cite{21} to align the powder samples. Since the easy
magnetization direction is along the $ab$ planes, the crystalline grains can be aligned with the $ab$ planes
parallel to the external field in a liquid medium by applying a high magnetic field. Upon cooling down under the
field the medium freezes, thus the grain orientation is fixed. In our experiment we chose the saturated NaCl
solution as the medium in order to suppress the ion exchange between hydroniums and Na$^{+}$ ions.
\cite{20,22,23}  Mixture of the powdered superconductor and saturated NaCl solution in a teflon container was
frozen below 250 K under an 8 T field. Magnetization measurement, performed on a Quantum Design PPMS facility,
showed that the aligned sample had relatively large magnetization compared with the non-aligned sample.

\section{\label{sec:level1}RESULTS AND DISCUSSION}

Fig. 1 shows the temperature dependence of ac magnetic susceptibility for
Na$_{0.33}$(H$_{3}$O)$_{0.02}$CoO$_{2}\cdot 1.4$H$_{2}$O sample with the field parallel to $ab$ planes. Sharp
superconducting transition can be seen at 4.5 K. The diamagnetic signal at 2 K is among the best results (for
polycrystalline samples) in the literatures, \cite{1,9,15} suggesting high quality of the present sample. The
inset shows an expanded plot for the normal-state susceptibility. One can see an obvious susceptibility upturn
with decreasing temperature. Similar phenomenon was previously reported.\cite{4,8}

\begin{figure}[tbp]
\includegraphics[width=8cm]{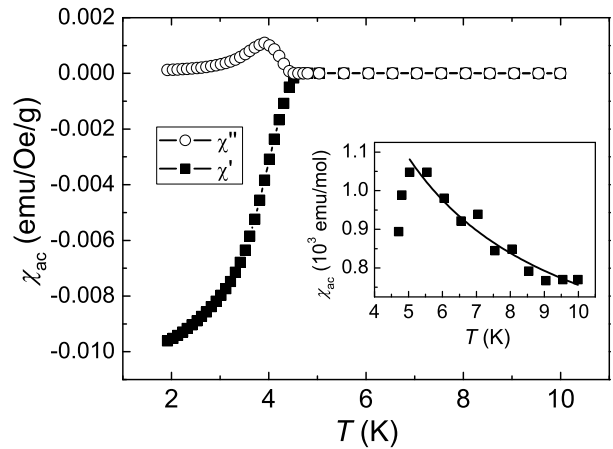}
\caption{Temperature dependence of ac magnetic susceptibility for aligned powder sample of
Na$_{0.33}$(H$_{3}$O)$_{0.02}$CoO$_{2}\cdot 1.4$H$_{2}$O with an ac field of 10 Oe parallel to $ab$ planes.
$\chi'$ and $\chi''$ denote the real and imaginary parts, respectively. The inset is an expanded plot for
displaying the normal-state susceptibility. The solid line is a Curie fit with the formula $\chi$=$\chi_{0}+C/T$
.}
\end{figure}

There exists at least two kinds of explanations about the susceptibility upturn behavior. One possibility is
that the susceptibility enhancement above $T_{c}$ is relevant to spin fluctuations.\cite{4,24,25} However, the
enhancement was sample dependent.\cite{26} Another explanation simply concerns the paramagnetic
impurity.\cite{4,21} The applied field lines up the impurity moments hence induces a magnetization, which can be
expressed by the following equation,

\begin{equation}
M=ng_{J}\mu_{B}JB_{J}(\alpha),
\end{equation}

where $n$ denotes the number of magnetic impurity per formula unit, $g_{J}$ the Lande $g$-factor, $J$ the
angular moment, $\mu_{B}$ the Bohr magneton,  $\alpha\equiv (g_{J}J\mu_{B}\mu_{0}H)/(k_{B}T)$, $\mu_{0}$ the
permeability in vacuum and $B_{J}(\alpha)$ is called Brillouin function,

\begin{equation}
B_{J}(\alpha)=\frac{2J+1}{2J}\coth(\frac{2J+1}{2J}\alpha)-\frac{1}{2J}\coth\frac{\alpha}{2J}.
\end{equation}

For a large $\alpha$ (at very low $T$ and/or very high $H$), $M$ tends to saturate. While for $\alpha$$\ll$1 (at
high $T$ and/or low $H$), the Brillouin function can be simplified and the susceptibility is given in the form
of Curie law,

\begin{equation}
\chi=\frac{M}{H}\approx\frac{n\mu_{0}\mu^{2}_{eff}}{3k_{B}T},
\end{equation}
where $\mu_{eff}$=$g_{J}\sqrt{J(J+1)}\mu_{B}$.

For the present study, the total molar magnetization above $T_{c}$ can be expressed as,

\begin{equation}
M_{tot}(H,T)=\chi_{0}H+M(H/T),
\end{equation}

where $\chi_{0}$ includes the contribution from Pauli paramagnetism as well as core diamagnetism, which is
independent of both temperature and magnetic field. Experimentally, $\chi_{0}$ value was obtained by the Curie
fit shown in the inset of Fig. 1. Therefore, ($M_{tot}-\chi_{0}H$) should be scaled in terms of $H/T$. As shown
in Fig. 2, the $T$-dependent part of normal-state $M(H,T)$ (0.1 T$\leq H\leq$8 T, 5 K$\leq T \leq$10 K) data
actually fall onto the same curve. Furthermore, equation (1) gives a satisfactory fit of the curve. The fitting
parameters are $n$=0.0014, $g_{J}$=2.4$\pm$0.1, $J$=1.6$\pm$0.2. We have also checked validity of the scaling
behavior on different samples. The $g_{J}$ and $J$ value are almost identical for all the samples. However, the
$n$ value varies from sample to sample, even for the samples with nearly the same $T_{c}$. This sample-dependent
behavior strongly suggests that the $T$-dependence part of the normal-state magnetization is not intrinsic.

\begin{figure}[tbp]
\includegraphics[width=8cm]{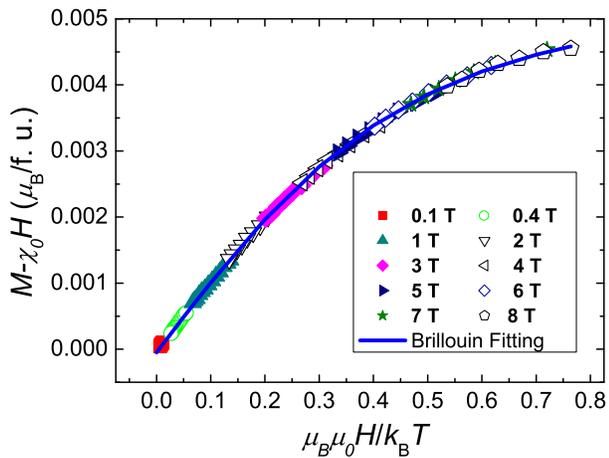}
\caption{(Color Online) Scaling behavior of the normal-state magnetization for the aligned
Na$_{0.33}$(H$_{3}$O)$_{0.02}$CoO$_{2}\cdot 1.4$H$_{2}$O superconductor.}
\end{figure}

As we know, Co$^{2+}$ in the high spin state has $J$=$S$=3/2 and $g$=2.5, which is very close to the fitting
result. Thus, the susceptibility upturn probably arises from the magnetic ion Co$^{2+}$ formed through charge
disproportionation reaction.\cite{23} Since the effective moments of Co$^{2+}$ is 4.8 $\mu_{B}$, according to
the Curie fit in the inset of Fig. 1, the Co$^{2+}$ concentration would be 0.13(2) \%, consistent with the above
Brillouin fitting.

Bearing it in mind that the $T$-dependent part of normal-state magnetization $M(H,T)$ can be well described by
the Brillouin function, we can subtract the magnetic impurity contribution from the total magnetization. After
this correction, the superconducting diamagnetism is linear with temperature near the transition region.
According to the criterion of the superconducting transition,\cite{27} the $T_{c}$($H$) values were determined
as the intersection of linear extrapolation of the magnetization with the base line, as shown in Fig. 3. The
result shows that $T_{c}$ is gradually suppressed with increasing applied field. It is noted that no
superconducting transition can be observed under 8 T down to the lowest temperature 1.9 K. This is in sharp
contrast with the previous magnetization measurements.\cite{8} At high fields and low temperatures, in fact, the
impurity moments tend to saturate, which could be mistaken for a superconducting transition.

\begin{figure}[tbp]
\includegraphics[width=8cm]{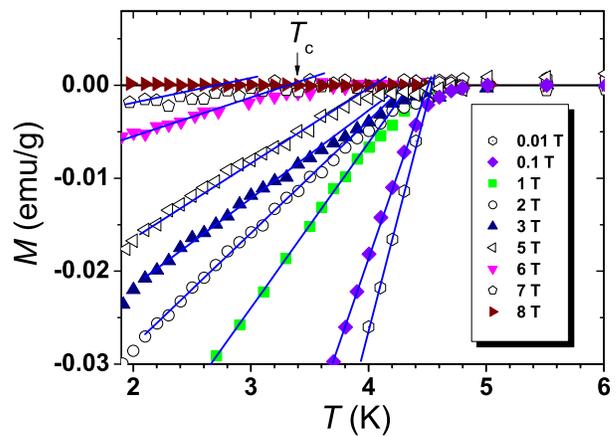}
\caption{(Color Online) Temperature dependence of the magnetization (after subtracting the contribution from
impurity magnetic moments) under different magnetic field for the aligned
Na$_{0.33}$(H$_{3}$O)$_{0.02}$CoO$_{2}\cdot 1.4$H$_{2}$O superconductor.}
\end{figure}

The $H^{\|ab}_{c2}-T$ phase diagram obtained is shown in Fig. 4. Our phase diagram is basically in agreement
with the measurements from resistivity,\cite{14} thermodynamic\cite{9,15} and $^{59}$Co-NMR.\cite{16} Although
the upper critical field at zero temperature $H^{\|ab}_{c2}$(0) could not be directly measured in present work,
the rapid decrease of $T_{c}$ at high magnetic field suggests that $H^{\|ab}_{c2}$(0) is much smaller than those
of the previous results by magnetic measurements.\cite{4,8} In fact, the specific-heat measurement shows no
superconducting transition down to 0.6 K at 8 T.\cite{15} That is to say, the $H^{\|ab}_{c2}$(0) value is very
close to the Pauli paramagnetic limit $H_{P}\sim$ 8.3 T, supporting spin singlet superconductivity in
Na$_{x}$(H$_{3}$O)$_{z}$CoO$_{2}\cdot y$H$_{2}$O. It is noted that recent $^{59}$Co Knight-shift
measurement\cite{28} also indicated that the electron pairing in the superconducting state is in the spin
singlet form.

\begin{figure}[tbp]
\includegraphics[width=8cm]{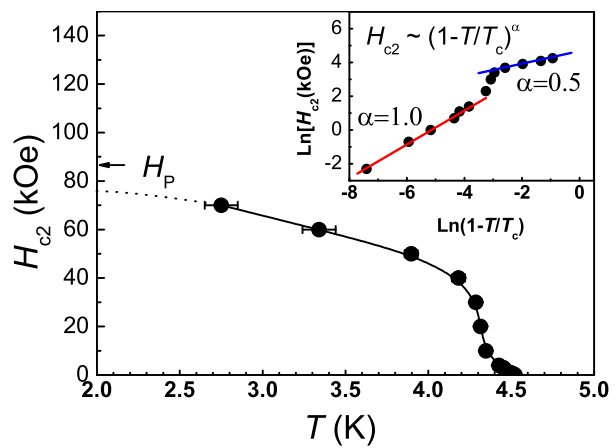}
\caption{Temperature dependence of upper critical field $H^{\|ab}_{c2}$($T$) for the aligned
Na$_{0.33}$(H$_{3}$O)$_{0.02}$CoO$_{2}\cdot 1.4$H$_{2}$O superconductor.}
\end{figure}

Fig. 4 also shows that an upward curvature or a slope change near $T_{c}$. Since our measurements were performed
on aligned samples, the effect of anisotropy of $H_{c2}$ can be ruled out. Therefore, the slope change of
$H^{\|ab}_{c2}$($T$) should be an intrinsic phenomenon.

There were theoretical explanations for the $H_{c2}$($T$) anomaly in terms of pairing symmetry
transition.\cite{8,19} Considering the two-dimensionality of the present superconductor, here we propose another
possibility: a field-induced 3D-to-2D dimensional crossover. As can be seen from the inset of Fig. 4,
$H^{\|ab}_{c2}(T)$ is proportional to ($1-T/T$$_{c}$) near $T_{c}$, as expected from 3D Ginzburg-Landau theory.
On the other hand, $H^{\|ab}_{c2}(T)$ is proportional to ($1-T/T_{c}$)$^{1/2}$ for $0.6T_{c} \leq T \leq
0.95T_{c}$, which is a signature of decoupled 2D behavior.\cite{29} Deutscher $et$ $al$.\cite{30} have shown
that a layered superconductor built up by alternating superconducting and insulating layers shows a dimensional
crossover characterized by an upturn in $H^{\|ab}_{c2}(T)$ at the temperature where interlayer coherence length
$\xi_{\bot}$($T$) is of the order of the sum of single superconducting and insulating layers. The $\xi_{c}$(0)
value of $\sim$ 13 \AA \cite{14} is comparable to the interlayer spacing 9.8 \AA, indicating that the
dimensional crossover is probable for the present superconductor. In order to clarify this issue, the
measurement of angular dependence of upper critical field on high quality single crystal is highly desired. As
is known, in the 3D regime $H^{\|ab}_{c2}(\Theta)$ shows a rounded maximum around $\Theta$=0 while in the 2D
case it exhibits a cusp.\cite{31}

\section{\label{sec:level4}CONCLUSION}

In summary, we have carried out a detailed magnetization measurement on aligned
Na$_{0.33}$(H$_{3}$O)$_{0.02}$CoO$_{2}\cdot 1.4$H$_{2}$O superconductors with magnetic field parallel to $ab$
planes. The temperature-dependent part of normal-state magnetization was revealed to arise from paramagnetic
impurity rather than spin fluctuations. The onset of superconductivity was characterized by the deviation from
normal-state magnetization governed by the Brillouin function. The result suggested that the $H^{\|ab}_{c2}$(0)
value for the cobalt oxyhydrate superconductor was just within the Pauli paramagnetic limit $H_{P}\approx$ 8.3
T, which supports spin-singlet superconductivity. Besides, the slope change near $T_{c}$ in $H^{\|ab}_{c2}-T$
phase diagram was confirmed to be intrinsic. A field-induced 3D-to-2D dimensional crossover is an alternative
explanation for this anomaly, which needs further investigations.

\begin{acknowledgments}
We acknowledge the supports from the National Basic Research Program of China (No.2006CB601003) and the National
Science Foundation of China (Project Nos.10674119 and 10634030).
\end{acknowledgments}

\end{document}